%% file: colm2025_conference.tex
\documentclass{article} 
\usepackage[preprint]{colm2025_conference}
\usepackage{algorithm}
\usepackage{algpseudocode}
\usepackage{algorithm}
\usepackage{algpseudocode}
\usepackage{amsmath}

\usepackage{graphicx}  
\usepackage{subcaption}  
\usepackage{tikz}  
\usepackage{pgfplots}  
\pgfplotsset{compat=1.17}  

\usepackage{microtype}
\usepackage{hyperref}
\usepackage{url}
\usepackage{booktabs}

\usepackage{lineno}

\definecolor{darkblue}{rgb}{0, 0, 0.5}
\hypersetup{colorlinks=true, citecolor=darkblue, linkcolor=darkblue, urlcolor=darkblue}

\title{FlowKV: A Disaggregated Inference Framework with Low-Latency KV Cache Transfer and Load-Aware Scheduling}

\author{
Weiqing Li$^*$, Guochao Jiang$^*$, Xiangyong Ding, Zhangcheng Tao, Chuzhan Hao, \\ \textbf{Chenfeng Xu, Yuewei Zhang, Hao Wang$^\dag$} \\
Alibaba Cloud Computing \\
\texttt{\{liweiqing.lwq, anyue.jgc\}@alibaba-inc.com} \\
\texttt{cashenry@126.com}
}

\begin{document}

\ifcolmsubmission
\linenumbers
\fi

\maketitle
\def\thefootnote{*}\footnotetext{Equal contributions.}
\def\thefootnote{$\dag$}\footnotetext{Corresponding author.}
\def\thefootnote{\arabic{footnote}}
\begin{abstract}
Disaggregated inference has become an essential framework that separates the prefill (P) and decode (D) stages in large language model inference to improve throughput. However, the KV cache transfer faces significant delays between prefill and decode nodes. The block-wise calling method and discontinuous KV cache memory allocation increase the number of calls to the transmission kernel. Additionally, existing frameworks often fix the roles of P and D nodes, leading to computational imbalances. In this paper, we propose FlowKV, a novel disaggregated inference framework, which reduces the average transmission latency of KV cache by 96\%, from 0.944s to 0.053s, almost eliminating the transfer time relative to the total request latency by optimizing the KV cache transfer. FlowKV introduces the Load-Aware Scheduler for balanced request scheduling and flexible PD node allocation. This design maximizes hardware resource utilization, achieving peak system throughput across various scenarios, including normal, computational imbalance, and extreme overload conditions. Experimental results demonstrate that FlowKV significantly accelerates inference by 15.2\%-48.9\% on LongBench dataset compared to the baseline and supports applications with heterogeneous GPUs.

\end{abstract}

\input{latex/introduction}
\input{latex/related_work}
\input{latex/methodology}

\input{latex/experiments}
\input{latex/conclusion}

\bibliography{colm2025_conference}
\bibliographystyle{colm2025_conference}

\input{latex/Appendix}

\end{document}

%% file: latex/introduction.tex
\section{Introduction }

Transformer-based \citep{transformer} Large Language Models (LLMs), such as GPT-4 \citep{gpt-4} and LLaMA-3 \citep{llama-3}, have become milestones in generative AI and provide new solutions for many industry scenarios. The vast potential of LLMs has reshaped many fields, including knowledge graphs \citep{knowledge-graph}, code generation \citep{code-generation}, role-playing agents \citep{role-playing}, and information extraction \citep{information-extraction, toner, mitigating}. The rapid growth in industrial demand has created new requirements and challenges for the end-to-end inference service performance of LLMs.

LLM inference has two stages: the compute-bound prefill stage and the memory-bound decode stage. Recent integrated inference frameworks like vLLM \citep{vllm} and SGLang \citep{sglang} run both stages in the same instance, which improves inference performance. However, these methods ignore the distinct computing needs of each stage, leading to interference between them and reduced service throughput. Recent studies \citep{splitwise, memserve, distserve, mooncake} focus on disaggregated prefill and decode nodes to optimize them independently for specific tasks. This disaggregated inference framework can also scale horizontally and work with different types of computing hardware, which helps improve overall service throughput and reduce hardware costs.

\begin{figure}[t]
    \centering
    \includegraphics[width=0.95\linewidth]{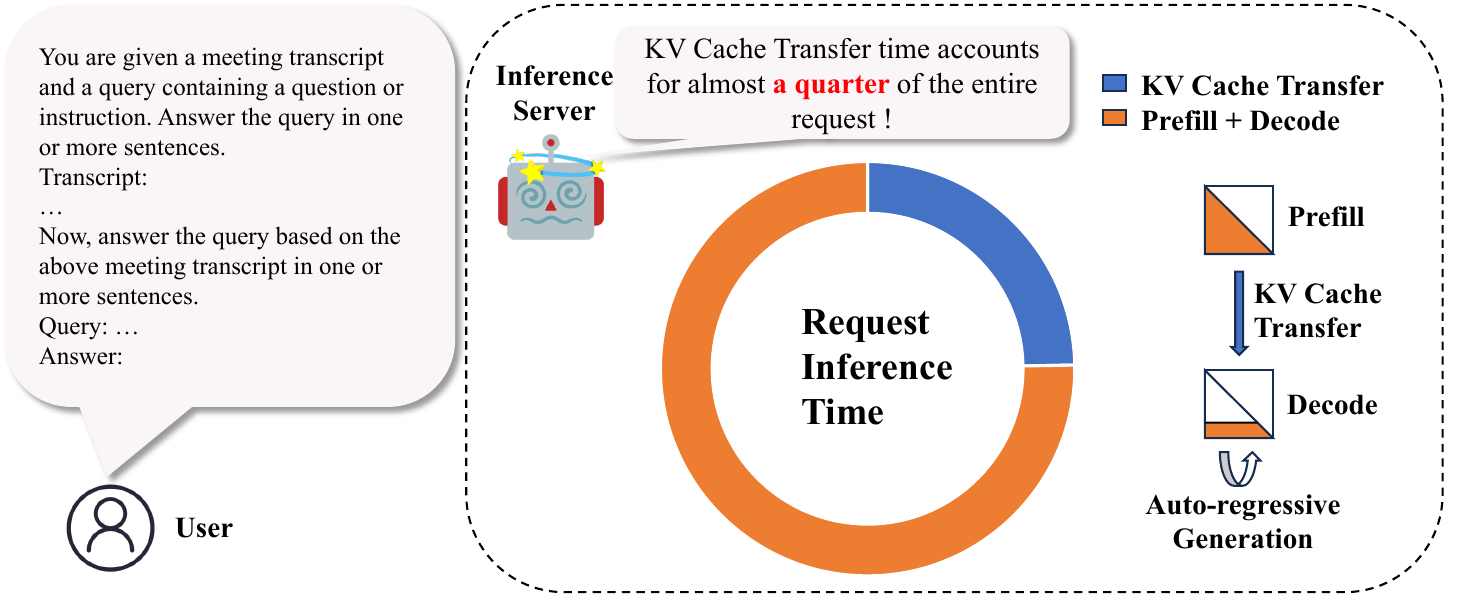}
    \caption{Time distribution of Prefill + Decode and KV Cache Transfer in a single request with NCCL-based transfer based original PagedAttention \citep{vllm}. The request is sampled from the LongBench \citep{longbench}, with an input length of 13k and an output length of 100. In this case, the KV Cache Transfer time between the prefill node and the decode node in the disaggregated framework accounts for about a quarter of the entire inference latency.}
    \label{fig:front}
\end{figure}

The disaggregated inference framework requires transmitting KV Cache between prefill nodes and decode nodes, an issue often overlooked in current research, as shown in Figure~\ref{fig:front}. Existing transmission schemes fall into two categories: RDMA-based transfer and NCCL-based transfer. Mooncake \citep{mooncake} uses an RDMA-based transfer solution that needs specific hardware. Its GPU VRAM exchange requires NVIDIA Mellanox NICs, and without them, we have to choose other solutions. Considering NCCL's good protocol compatibility (RoCE, IB, Socket), many studies have chosen to utilize NCCL. Splitwise \citep{splitwise} uses an NCCL-based solution that transfers data at the LLM layer level. While this approach overlaps communication and computation, its effectiveness is limited due to frequent NCCL transfers. vLLM-disaggregated \citep{vllm} uses NCCL-based buffer transfers, combining disaggregated KV Cache layer by layer into a continuous buffer before transmission. However, merging discrete tensors into complete ones causes extra memory usage and time delays, reducing overall throughput. 

To address these challenges, we propose a framework that optimizes both the KV Cache structure and memory block allocator to minimize transfer time. While standard PagedAttention \citep{vllm} manages memory at block level and requires inter-node communication at the same level for KV Cache Transfer. Our KV Cache structure adjustment method reduces the number of NCCL communications, providing two benefits. First, fewer transmissions reduce the total KV Cache Transfer time with fewer communication kernels. Second, fewer communications reduce GPU computing task blocking time, which lowers overall request latency. In addition, we also propose a Load-Aware Scheduler designed to enhance the overall throughput of the inference service.

Our main contributions in this paper are summarized as follows:

\begin{itemize}

\item We analyze the communication patterns between prefill and decode nodes in the disaggregated inference framework. Based on our analysis, we propose modifications to the KV Cache structure and its allocator to reduce communication overhead. These optimizations effectively eliminate the KV Cache transfer time relative to the total request latency.

\item We implement Load-Aware Scheduling between prefill and decode nodes, which significantly improves the overall throughput of our inference service.

\item Through extensive experiments, we demonstrate the superior performance of our framework compared to existing open-source inference frameworks and validate its effectiveness across heterogeneous GPU environments.

\end{itemize}

%% file: latex/related_work.tex
\section{Related Work}

\textbf{Inference Serving.} More and more inference serving systems are being proposed and updated to cope with the rapidly developing LLM application requirements. Orca \citep{orca} introduces continuous batching to improve the overall service throughput. vLLM \citep{vllm} uses paged attention to manage KV Cache memory, further alleviating the problem of GPU memory fragmentation. Sarathi \citep{sarathi} introduces chunked prefill to allow the service to perform prefill and decode requests simultaneously, optimizing the throughput of the overall inference service. TorchServe \footnote{\url{https://pytorch.org/serve/}} and NVIDIA Triton \footnote{\url{https://developer.nvidia.com/triton-inference-server}} work on inference serving optimization for common transformer models. These inference serving frameworks generally process the prefill and decode tasks on the same node instance, and fail to consider the differences between the two-stage tasks compared to the disaggregated inference serving framework.

\noindent\textbf{Prefill/Decode Disaggregated Inference.} In light of the distinct task characteristics of LLM in the prefill and decode stages, several studies \citep{splitwise, distserve, memserve, mooncake} have explored Prefill/Decode Disaggregated Inference, achieving some improvements. All previous works separate prefill and decoding instances to enhance inference performance and propose a global scheduler for request distribution. Splitwise \citep{splitwise} maintains a mixed pool for expanding contracts as needed by the workload and uses a hierarchical two-level scheduling system for managing the pool and requests. MemServe \citep{memserve} supports co-located P/D instances, in addition to separate prefill and decoding ones, with each instance having a memory pool for allocation, index management, and distributed transfer. Mooncake \citep{mooncake} features a KVCache-centric architecture that separates prefill and decoding clusters. However, these studies often overlook the latency of KV Cache transfer in single-request scenarios and thus fail to optimize this aspect.

%% file: latex/methodology.tex
\section{Methodology }


In this section, we first introduce the background of the disaggregated inference framework and then present the innovative optimizations of FlowKV framework proposed in this paper.

\subsection{Background: Prefill/Decode Disaggregated Inference}

Formally, modern LLM inference framework employ autoregressive generation \citep{llm_survey}, iteratively predicting the next token $y_{t+1}$ based on the $t$ input tokens $x = \{x_1, x_2, \cdots, x_t\}$, until encountering the end-of-sequence (EOS) token.

The generation process consists of two distinct phases. 
\begin{itemize}
    \item \textbf{Prefill (P)}: Computing the initial token $ y_{t+1} $ and the associated KV cache of all $t+1$ tokens.
    \item \textbf{Decode (D)}: Iteratively generating subsequent tokens $ \{y_{t+2}, y_{t+3}, \cdots, y_{t+k}\} $ by processing only the newly generated token from the previous step while maintaining accumulated KV cache through incremental updates.
\end{itemize}
Most existing LLM inference systems \citep{vllm, sglang} adopt a prefill-decode-colocated (PD-colocated) framework, where both phases are co-located on the same GPU devices despite their distinct computational characteristics. The Prefill/Decode disaggregated inference (PD-disaggregated) framework places the P nodes and D nodes on different devices, avoiding interference between the P and D phases. 

Let $\mathcal{P} = \{P_i\}_{i=1}^N$ and $\mathcal{D} = \{D_j\}_{j=1}^M$ represent the clusters of prefill and decode nodes, respectively, where $N$ is the number of P nodes and $M$ is the number of D nodes.  First, when the inference system receives a request $R$ with the input token sequence $x = \{x_1, x_2, \cdots, x_t\}$, it selects the $p$-th target prefill node $P_p$ and forwards request $R$ to $P_p$ for prefill computation to generate the first token $y_{t+1}$ and KV Cache tensors $\mathbf{K} = \{ \mathbf{k}_1, \mathbf{k}_2, \cdots, \mathbf{k}_t, \mathbf{k}_{t+1} \}$ and $\mathbf{V} = \{ \mathbf{v}_1, \mathbf{v}_2, \cdots, \mathbf{v}_t, \mathbf{v}_{t+1} \}$, where $\mathbf{k}_i$ and $\mathbf{v}_i$ are $i$-th token's KV Cache tensors. The process can be defined as follows:
\begin{align}
    y_{t+1}, \mathbf{K}, \mathbf{V} = P_p(R).
\end{align}

Second, after request $R$ is processed, the inference system selects $d$-th decode node $ D_d $ and quickly transfers the KV Cache $\mathbf{K}$ and $\mathbf{V}$ from node $P_p$ to node $D_d$. 


Third, the inference system forwards request $R$ to $D_d$ to continue generating the next tokens $\{y_{t+1}, y_{t+2}, \cdots, y_{t+k}\}$ iteratively. We define the request $R$ forwarded to node $D_d$  for decoding computation and sampling next token $y_{t+i} (1 < i \le k)$ as follows:
\begin{align}
    y_{t+i}, \mathbf{k}_{t+i}, \mathbf{v}_{t+i} &= D_d\left( \{x,y_{t+1},\cdots, y_{t+i-1}\}, \{ \mathbf{k}_1, \cdots,\mathbf{k}_{t+i-1} \}, \{ \mathbf{v}_1, \cdots,\mathbf{v}_{t+i-1} \} \right), \\
    \mathbf{K} &= \{ \mathbf{k}_1, \mathbf{k}_2, \cdots, \mathbf{k}_{t+i} \}, \\
    \mathbf{V} &= \{ \mathbf{v}_1, \mathbf{v}_2, \cdots, \mathbf{v}_{t+i} \}.
\end{align}

Obviously, PD-disaggregated faces two challenges: 1. Performing a low-latency transfer of the KV cache between PD nodes, especially between heterogeneous nodes. 2. Maintaining the load balance between the P and D nodes.

\subsection{Overview of the Main Framework of FlowKV}







\begin{figure}
    \centering
    \includegraphics[width=1\linewidth]{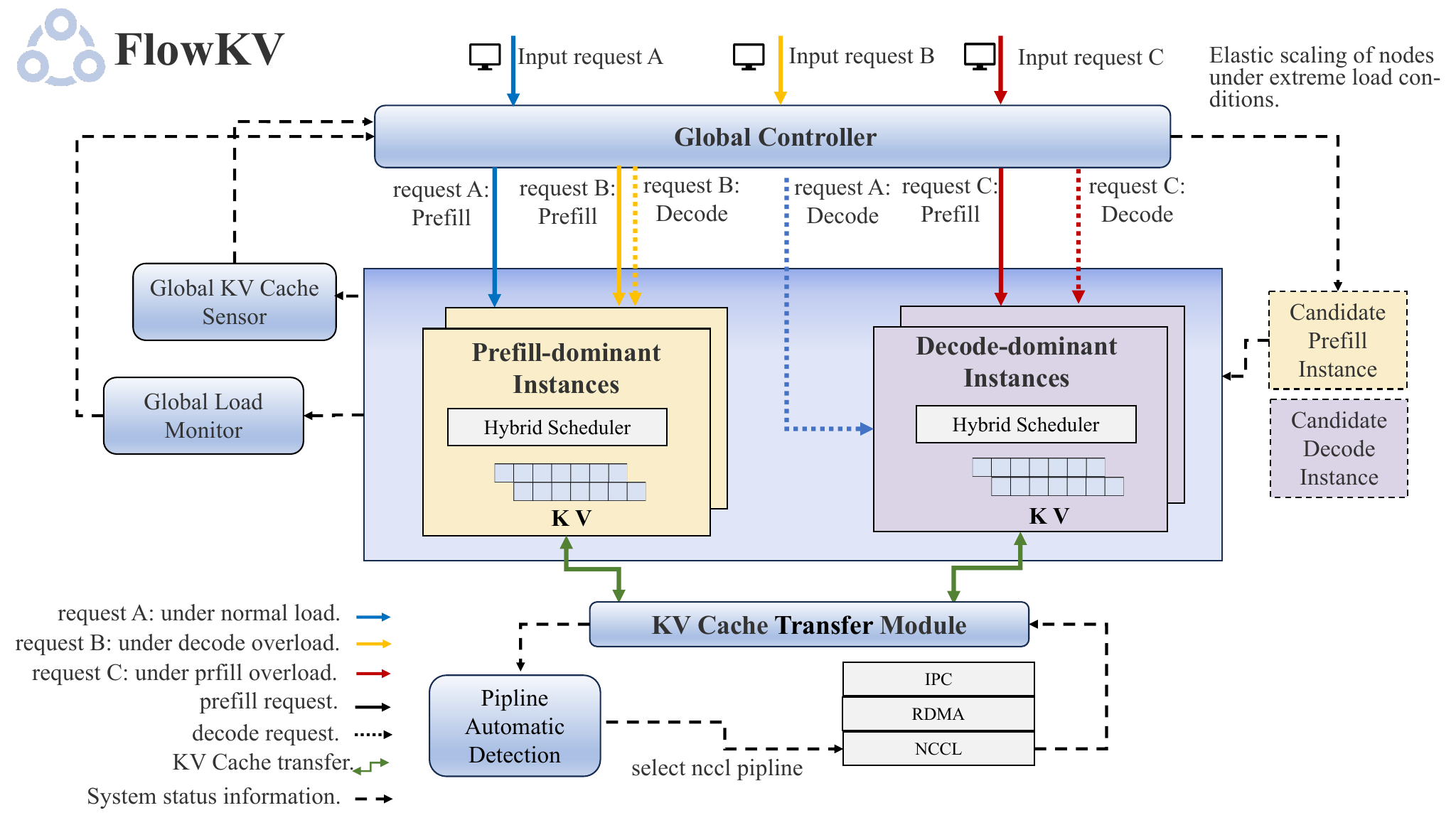}
    \caption{FlowKV framework. FlowKV enables high-speed transmission of KV Cache between P and D nodes through KV Cache transfer module. Furthermore, it employs a global controller to monitor the workload and KV cache status of P and D nodes in real-time. Under different load conditions (normal load, imbalanced load, extreme load), the system adopts load balancing strategies for request scheduling or performs elastic node scaling to ensure efficient resource utilization.}
    \label{fig:Figure2}
\end{figure}

We propose the FlowKV framework, which effectively cuts down the cost and latency of KV Cache transfer. It introduces a Load-Aware Scheduler to keep all nodes running in a balanced state. FlowKV includes five main modules: Prefill nodes, Decode nodes, a global controller, hybrid schedulers, and a KV Cache transfer module, as shown in Figure \ref{fig:Figure2}.

FlowKV separates the dependencies between P nodes and D nodes, allowing them to run on any nodes regardless of number, ratio, or machine architecture. The global controller is FlowKV's central component, managing the scheduling of P/D processes and requests. It monitors load patterns and identifies global cache prefix matches to boost throughput and reduce KV Cache transfer latency. By creating optimal request scheduling schemes, the global controller ensures all nodes in the cluster work efficiently and supports elastic scaling of P and D nodes during overloads. The KV Cache transfer module, another key part of FlowKV, selects the best transfer pipeline based on hardware features. Unlike other architectures, FlowKV's P and D nodes perform hybrid computation when there's a computational imbalance. This capability is implemented by the hybrid scheduler within each P and D node, in collaboration with the global controller.

\subsection{A cost-effective low-latency KV Cache transfer solution
}

To handle different hardware setups, we've improved the KV Cache transfer between nodes, supporting various KV transfer methods like NCCL, Inter-Process Communication (IPC), and Remote Direct Memory Access (RDMA). In single-machine setups, IPC is the default method. In mixed deployments, NCCL is preferred because of its great performance and popularity for inter-node GPU communication.

While NCCL performs well, it only supports data transfers between contiguous memory addresses \citep{KVDirect, memserve}. This limitation becomes noticeable in disaggregated inference scenarios. Modern inference frameworks typically use PagedAttention \citep{vllm} to manage the KV Cache, improving memory use through logical-to-physical block mapping. However, fragmented physical blocks require NCCL to handle many small data chunks during cross-GPU transfers. This leads to a sharp increase in NCCL API calls and transmission latency. Additionally, frequent NCCL kernel launches may compete for SM resources with processes like GEMM, potentially causing more inference latency.

FlowKV introduces specific optimizations that cut down the latency of KV Cache transfer in disaggregated inference. By aiming to minimize how often the KV Cache is transfered, it first tweaks the KV Cache shape managed by PagedAttention. This change combines the separate tensor structures across all layer dimensions into a continuous tensor, reducing the number of NCCL API calls. The KV Cache shape transformation can be expressed as
\begin{equation}
\mathbf{K}, \mathbf{V}: (L, 2, B, H) \rightarrow (B, L, 2, H),
\end{equation}
where $B$ is the number of blocks, $L$ represents the number of model layers, $H$ represents the dimension of the remaining KV Cache vector. Thus, the number of NCCL API calls per KV block is reduced by a factor of \textit{$L$ $\times$ 2}. Additionally, targeted optimizations are made for the PagedAttention kernel.

Second, FlowKV uses segment management techniques from operating systems to improve how traditional block allocators assign and release memory. It tries to allocate new requests within one or a few contiguous segments. It maintains continuous KV Cache blocks within these segments. FlowKV manages free memory blocks with segment-based minimum heaps. It chooses the right segments during allocation to minimize waste and merges adjacent free segments during deallocation to boost future allocation efficiency.

Finally, for requests that require KV Cache transfer, before triggering NCCL API calls, FlowKV performs \textit{bidirectional segment alignment} on the KV Cache block ID lists. After alignment, the output results determine which block IDs can be merged for a single send-receive operation via NCCL, thereby minimizing the number of transfers, as illustrated in Figure \ref{fig:figure3} in the Appendix.



\subsection{Load-Aware Scheduler for optimal compute power utilization}

For PD-disaggregated inference systems, using a fixed ratio of P and D nodes isn't flexible. Under heavy loads and varying request patterns, this fixed ratio can easily overload one type of nodes, causing computational resource issues \citep{distserve}. To address this, we propose a Load-Aware Scheduling scheme. It uses a global controller and local schedulers in each P and D node to monitor each node's load status in real-time. Additionally, it balances the routing for prefill requests and decode requests through communication between the global controller and the local schedulers.

The global controller's main job is to create request scheduling plans based on each node's workload and the system's current load. It determines the best nodes to handle incoming requests. Local schedulers, or hybrid schedulers, manage both a prefill scheduler and a decode scheduler. Like vLLM's scheduler, each one has separate running, waiting, swapped, and pending queues, along with specific logic. They share a block manager with the hybrid scheduler. The hybrid scheduler manages the inference process by coordinating the prefill and decode schedulers. During each scheduling cycle, it can prioritize sub-schedulers based on the global controller's instructions. By default, prefill has priority, so all nodes focus on prefill requests when they are available.

To describe the current state of the inference system, we define three load scenarios: normal load, imbalanced load, and extreme load. Imbalanced load happens when either P or D nodes get more requests than they can handle. This causes one type of node to be GPU compute-bound while the other has low GPU use. Extreme scenarios include system overload and low load conditions. System overload occurs when both P and D nodes get more requests than they can handle. A low load scenario means there is minimal demand. Each load scenario needs a different request scheduling strategy. We outline the application of the Load-Aware Scheduling scheme in normal load, computational imbalance, and extreme scenarios in the Appendix \ref{Load-Aware-Application}, and provide the pseudocode for Algorithm \ref{alg:Load-Aware}.

%% file: latex/experiments.tex
\tikzset{every picture/.style={execute at begin picture=\centering}}

\section{Experiments}

In this section, we compare the performance of our approach under both single-node homogeneous scenarios and multi-node heterogeneous scenarios against PD-colocate inference framework and existing open-source disaggregated inference systems. Additionally, we conduct performance comparison experiments for different KV Cache transfer pipelines.







\subsection{Datasets and Models}

\begin{itemize}
    \item \textbf{Simulated Data}: We generated a batch of simulated data with predefined input and output lengths to compare the maximum throughput of the systems.
    \item \textbf{Real-World Data}: We selected the summarization task as a real-world benchmark to compare end-to-end response latency (E2E). The data is sampled from the summarization tasks in the LongBench \citep{longbench} dataset, we select three subtasks: \verb|gov_report|, \verb|multi_news|, and \verb|qmsum|.
\end{itemize}

For all scenarios, we simulate requests using a Poisson arrival process and control the request rate via requests per second (RPS).

We selected the LLaMA-3 \citep{llama-3} series models (Meta-Llama-3.1-8B-Instruct \footnote{\url{https://huggingface.co/meta-llama/Llama-3.1-8B-Instruct}}, Meta-Llama-3.1-70B-Instruct \footnote{\url{https://huggingface.co/meta-llama/Llama-3.1-70B-Instruct}}) for testing, which are representative LLM series widely used in academia and industry.

\noindent\textbf{Baselines}. vLLM \citep{vllm} is a community-driven project contributed by both academia and industry. It uses PagedAttention to efficiently manage memory for attention keys and values combined with continuous batching to provide superior throughput performance. In addition to PD-colocated, vLLM currently also supports PD-disaggregated deployment. Other representative open-source PD-disaggregated inference systems: Mooncake \citep{mooncake}, DistServe \citep{distserve}.

\begin{itemize}
    \item \textbf{Homogeneous Performance Evaluation}: This evaluation is run on an NVIDIA A100-SXM4-80GB server with 8 GPUs interconnected via NVLink links.
    \item \textbf{Heterogeneous Performance Evaluation}: This evaluation is run on two servers, L20 and H20, equipped with 4 GPUs with 48GB memory and 8 GPUs with 96GB memory, respectively. The two servers are connected via Elastic Network Interfaces (ENI), providing a basic network bandwidth  allowing for low-latency KV Cache transfer.
\end{itemize}



\subsection{Homogeneous Deployment Scenario}

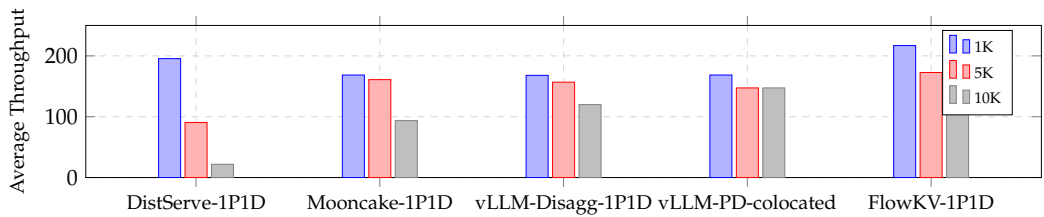
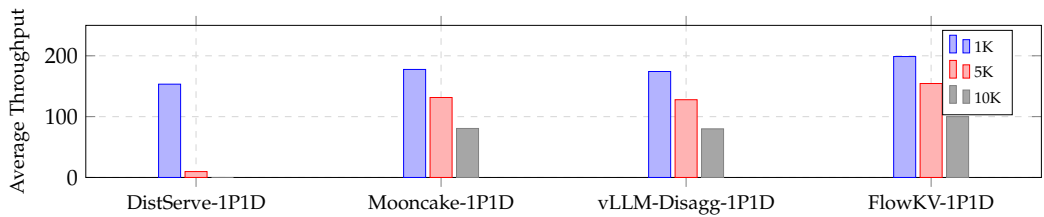
\begin{figure}[!ht]
    \centering
    \begin{subfigure}[b]{1\linewidth}  
        \centering
        \resizebox{1\linewidth}{!}{  
            \begin{tikzpicture}
                \begin{axis}[
                    ybar, 
                    bar width=0.35cm,  
                    symbolic x coords={DistServe-1P1D, Mooncake-1P1D, vLLM-Disagg-1P1D, vLLM-PD-colocated, FlowKV-1P1D},
                    xtick=data,
                    xticklabel style={font=\footnotesize},  
                    ymin=0, ymax=250,
                    ylabel={Average Throughput},
                    ylabel style={font=\footnotesize},
                    ytick style={font=\footnotesize},
                    legend pos=north east,
                    legend style={font=\scriptsize, cells={anchor=west}},
                    legend cell align={left},
                    width=1.2\textwidth,
                    height=4cm,
                    grid=major,
                    grid style={dashed, line width=0.5pt, draw=gray!30},
                    enlarge x limits=0.15,  
                    nodes near coords,
                    point meta=explicit symbolic,
                    every node near coord/.style={font=\footnotesize, anchor=west},
                ]
                    \addplot+[
                        ybar, fill=blue!30, draw=blue, nodes near coords,
                    ] 
                    coordinates {(DistServe-1P1D,195.45) (Mooncake-1P1D,168.49) (vLLM-Disagg-1P1D,167.99) (vLLM-PD-colocated,168.51)(FlowKV-1P1D,216.90)};
                    
                    \addplot+[
                        ybar, fill=red!30, draw=red, nodes near coords,
                    ] 
                    coordinates {(DistServe-1P1D,90.47) (Mooncake-1P1D,160.92) (vLLM-Disagg-1P1D,156.74) (vLLM-PD-colocated,147.31)(FlowKV-1P1D,172.63)};
                    
                    \addplot+[
                        ybar, fill= gray!50, draw=gray, nodes near coords,
                    ] 
                    coordinates {(DistServe-1P1D,21.81) (Mooncake-1P1D,93.57) (vLLM-Disagg-1P1D,119.93) (vLLM-PD-colocated,147.31)(FlowKV-1P1D,172.63)};
                    
                    \legend{1K,5K,10K}
                \end{axis}
            \end{tikzpicture}
        }
        \subcaption{Two nodes of Llama-3.1-8B-Instruct.}
        \label{fig:Figure4a}
    \end{subfigure}%
    \hspace{0.02\textwidth}  
    \begin{subfigure}[b]{1\linewidth}  
        \resizebox{1\linewidth}{!}{  
            \begin{tikzpicture}
                \centering
                \begin{axis}[
                    ybar, 
                    bar width=0.35cm,  
                    symbolic x coords={DistServe-1P1D, Mooncake-1P1D, vLLM-Disagg-1P1D, FlowKV-1P1D},
                    xtick=data,
                    xticklabel style={font=\footnotesize},  
                    ymin=0, ymax=250,
                    ylabel={Average Throughput},
                    ylabel style={font=\footnotesize},
                    ytick style={font=\footnotesize},
                    legend pos=north east,
                    legend style={font=\scriptsize, cells={anchor=west}},
                    legend cell align={left},
                    width=1.2\textwidth,
                    height=4cm,
                    grid=major,
                    grid style={dashed, line width=0.5pt, draw=gray!30},
                    enlarge x limits=0.15,  
                    nodes near coords,
                    point meta=explicit symbolic,
                    every node near coord/.style={font=\footnotesize, anchor=west},
                ]
                    \addplot+[
                        ybar, fill=blue!30, draw=blue, nodes near coords,
                    ] 
                    coordinates {(DistServe-1P1D,153.35) (Mooncake-1P1D,177.55) (vLLM-Disagg-1P1D,174.1825) (FlowKV-1P1D,198.7675)};
                    
                    \addplot+[
                        ybar, fill=red!30, draw=red, nodes near coords,
                    ] 
                    coordinates {(DistServe-1P1D,9.68) (Mooncake-1P1D,131.44) (vLLM-Disagg-1P1D,127.79) (FlowKV-1P1D,154.42)};
                    
                    \addplot+[
                        ybar, fill= gray!70, draw=gray, nodes near coords,
                    ] 
                    coordinates {(DistServe-1P1D,0) (Mooncake-1P1D,80.55) (vLLM-Disagg-1P1D,79.91)(FlowKV-1P1D,100.02)};
                    
                    \legend{1K,5K,10K}
                \end{axis}
            \end{tikzpicture}
        }
        \subcaption{Two nodes of Llama-3.1-70B-Instruct.}
        \label{fig:Figure4b}
    \end{subfigure}
    
    \caption{Throughput performance using simulated data in homogeneous deployment scenario}
    \label{fig:ThroughputFigure}
\end{figure}

In this section, we evaluate the maximum throughput performance of FlowKV under a single-node deployment using simulated data. In this setup, the KV Cache transfer method automatically switches to IPC mode to optimize performance. The simulated dataset consists of 100 entries, with input tokens varying from 1K, 5K, to 10K, while the output tokens are fixed at 256. We gradually increase the RPS and benchmark against DistServe, Mooncake, and vLLM. Figure \ref{fig:Figure4a} and Table \ref{table:1} illustrates the throughput performance of the Llama-3.1-8B-Instruct inference service powered by two GPUs. Compared to vLLM PD-colocated inference services, FlowKV significantly reduces mutual interference by disaggregating the P and D job processes. Additionally, this approach is enhanced by a Load-Aware Scheduling algorithm, which contributes to a 25\% increase in average throughput for input tokens of 1K, 5K, and 10K. Compared to other disaggregated inference frameworks configured with a 1P1D setup, such as DistServe, Mooncake, and vLLM-Disagg, FlowKV improves throughput by 95\%, 40\%, and 35\%, respectively.

\begin{table*}[t]
    \centering
    \resizebox{0.8\textwidth}{!}{\begin{tabular}{p{2cm}|p{1cm}|p{1.5cm}p{1.5cm}p{2cm}p{1.5cm}|p{1.5cm}}
    \toprule
    \toprule
    \bf Input/Output tokens& \bf RPS & \bf DistServe\newline \bf 1P1D & \bf Mooncake\newline \bf 1P1D & \bf vLLM-Disagg\newline \bf 1P1D & \bf vLLM\newline \bf PD-colocated & \bf FlowKV\newline \bf 1P1D \\
    \midrule
          1K/256&0.1& 27.62&  20.93& 20.72& 20.93&\bf 27.88\\ 
               &0.2& 54.78& 41.87& 41.84& 41.87&\bf 55.52\\ 
               &0.4& 107.45& 83.74& 83.74& 83.74&\bf 109.86\\ 
               &0.6& 158.17& 126.93& 125.12& 125.03&\bf 163.45\\ 
               &0.8& 207.70& 166.26& 167.49& 165.75&\bf 216.08\\
               &1.0& 254.04& 206.38& 208.00& 208.08&\bf 267.61\\ 
               &1.5& 349.36& 304.70& 302.97& 304.98&\bf 387.47\\
               &2.0& 404.55& 397.18& 394.05& 397.70&\bf 507.36 \\
    \midrule
              5K/256&0.1& 26.54&  21.08& 21.08& 21.07&\bf 27.86\\ 
               &0.2& 51.26& 42.17& 42.17& 42.17&\bf 55.48\\ 
               &0.4& 86.99& 83.00& 83.48& 83.40&\bf 109.91\\ 
               &0.6& 107.43& 123.73& 123.52& 124.11&\bf 163.3\\ 
               &0.8& 109.11& 165.41& 162.28& 163.80&\bf 215.35\\
               &1.0& 115.63& 204.65& 202.87& 203.04&\bf 264.22\\ 
               &1.5& 114.00& 291.05& 287.40& 294.10&\bf 382.34\\
               &2.0& 112.87& 356.28& 331.12& 378.97&\bf 470.68 \\
    \midrule
 10K/256& 0.1& 20.92& 20.52& 20.52& 20.51&\bf 27.85\\
 & 0.2& 21.94& 41.04& 41.06& 41.05&\bf 55.34\\
 & 0.4& 21.93& 81.89& 81.83& 82.05&\bf 109.44\\
 & 0.6& 21.92& 121.01& 120.82& 121.25&\bf 161.16\\
 & 0.8& 21.94& 155.78& 154.03& 161.02&\bf 208.61\\
 & 1.0& 21.94& 171.07& 171.11& 194.24&\bf 251.55\\
 & 1.5& 21.94& 63.71& 184.61& 271.35&\bf 281.93\\
 & 2.0& 21.94& Failure& 185.47& \bf286.97&285.14 \\
    \bottomrule
    \bottomrule
    \end{tabular}}
        \caption{Throughput comparison based on Llama-3.1-8B-Instruct. We conducted evaluations using synthetic requests with varying input token counts of 1K, 5K, and 10K, while maintaining a fixed output token count of 256.}
 \label{table:1}
\end{table*}

\begin{table*}[t]
    \centering
    \begin{tabular}{p{2cm}|p{1cm}|p{1.5cm}p{2cm}p{1.5cm}|p{1.5cm}}
    \toprule
    \toprule
    \bf Input/Output tokens& \bf RPS & \bf DistServe\newline \bf 1P1D & \bf Mooncake\newline \bf 1P1D & \bf vLLM-Disagg\newline \bf 1P1D & \bf FlowKV\newline \bf 1P1D \\
    \midrule
    1K/256 & 0.1 & 27.29 & 23.44 & 23.44 & \bf27.67 \\ 
       & 0.2 & 53.38 & 46.88 & 46.87 & \bf54.69 \\ 
       & 0.4 & 102.36 & 92.55 & 91.77 & \bf106.74 \\ 
       & 0.6 & 147.27 & 136.79 & 134.87 &\bf 156.28 \\ 
       & 0.8 & 188.57 & 179.65 & 176.80 &\bf 203.51 \\
       & 1.0 & 222.84 & 221.42 & 216.44 & \bf248.03 \\ 
       & 1.5 & 238.47 & 316.08 & 310.77 & \bf351.10 \\
       & 2.0 & 246.62 & 403.62 & 392.50 &\bf 442.12 \\
    \midrule
    5K/256 & 0.1 & 9.68 & 23.57 & 23.56 & \bf27.65 \\ 
       & 0.2 & 9.69 & 47.12 & 46.91 & \bf54.60 \\ 
       & 0.4 & 9.69 & 92.58 & 91.59 & \bf106.40 \\ 
       & 0.6 & 9.69 & 134.74 & 133.69 & \bf154.65 \\ 
       & 0.8 & 9.69 & 172.68 & 167.79 & \bf197.83 \\
       & 1.0 & 9.69 & 182.81 & 178.95 & \bf233.93 \\ 
       & 1.5 & 9.69 & 195.32 & 188.00 & \bf228.44 \\
       & 2.0 & 9.69 & 202.72 & 191.83 & \bf231.87 \\
    \midrule
    10K/256 & 0.1 & Failure & 23.49 & 23.49 & \bf27.62 \\
       & 0.2 & Failure & 46.67 & 46.62 & \bf54.41 \\
       & 0.4 & Failure & 86.46 & 85.88 & \bf103.47 \\
       & 0.6 & Failure & 97.10 & 96.14 & \bf121.93 \\
       & 0.8 & Failure & 97.45 & 96.70 &\bf 122.62 \\
       & 1.0 & Failure & 97.69 & 96.93 & \bf123.29 \\
       & 1.5 & Failure & 97.82 & 96.75 & \bf123.45 \\
       & 2.0 & Failure & 97.75 & 96.79 & \bf123.43 \\
    \bottomrule
    \bottomrule
    \end{tabular}
        \caption{Throughput comparison of Llama-3.1-70B-Instruct. DistServe encountered a failure when processing an input tokens of 10K.}
        \label{table:2}
\end{table*}

Additionally, we conducted a performance evaluation using a large-scale parameter model for comparative analysis. Figure \ref{fig:Figure4b} and Table \ref{table:2} illustrates the throughput performance of the Llama-3.1-70B-Instruct service deployed across eight A100 GPUs, configured as two nodes with intra-node tensor parallelism set to four. In comparison to DistServe, Mooncake, and vLLM-Disagg, FlowKV demonstrates substantial throughput enhancements, achieving improvements of 95\%, 39\%, and 35\%, respectively.

\subsection{Heterogeneous Deployment Scenario}
In this section, we conduct a comparative analysis of FlowKV and other baselines under heterogeneous deployment configurations, and evaluate E2E across practical application datasets. The key advantage of disaggregated heterogeneous inference lies in its capability to optimize service deployment by aligning GPU characteristics (e.g., memory bandwidth and capacity) with task-specific computational requirements. For summarization tasks demanding high decoding throughput, allocating decode operations to GPUs with superior memory resources (H20 nodes) achieves lower time-per-output-token (TPOT) and E2E.

\begin{figure}
  \centering
  \begin{subfigure}[t]{0.3\textwidth}
    \includegraphics[width=\linewidth]{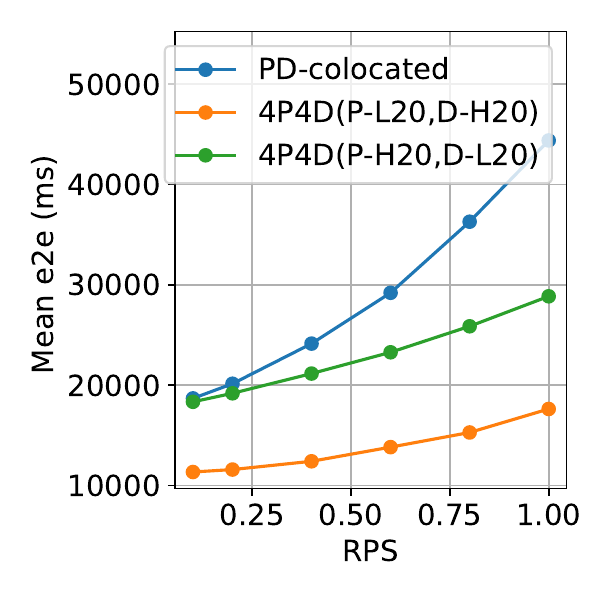}  
    \caption{E2E latency of Gov-report } 
    \label{fig:e2e_gr}
  \end{subfigure}
  \hfill
  \begin{subfigure}[t]{0.3\textwidth}
    \includegraphics[width=\linewidth]{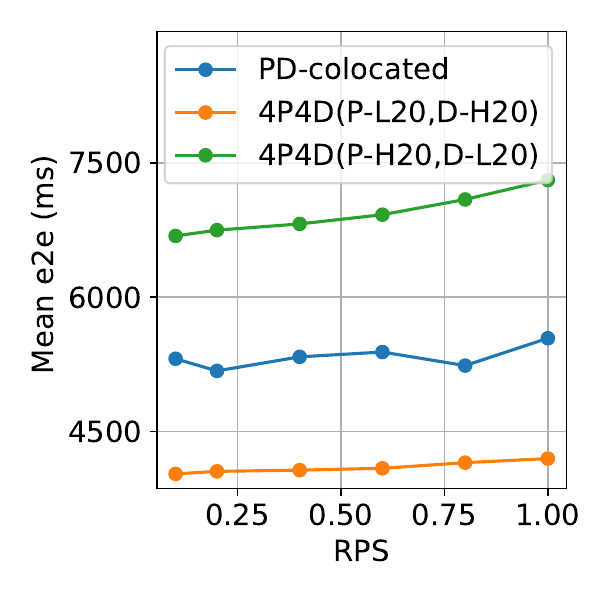} 
    \caption{E2E latency of Multi-news } 
    \label{fig:e2e_gr1}
  \end{subfigure}
  \hfill
  \begin{subfigure}[t]{0.3\textwidth}
    \includegraphics[width=\linewidth]{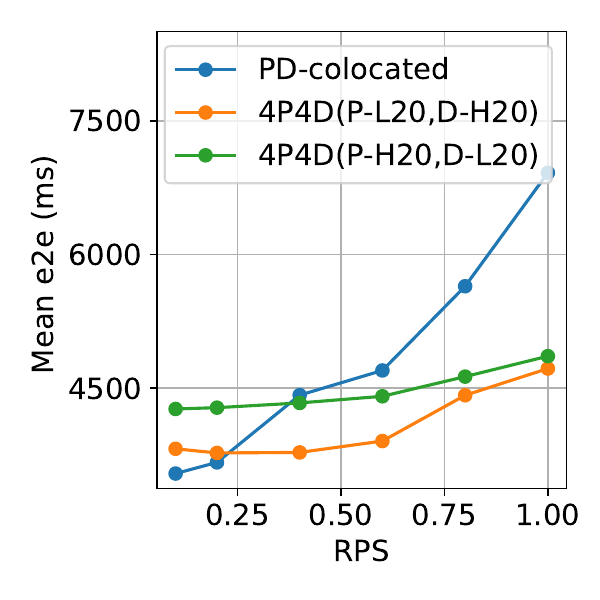} 
    \caption{E2E latency of Qmsum} 
    \label{fig:e2e_g3}
  \end{subfigure}
  \caption{E2E performance in heterogeneous deployment scenario} 
    \label{fig:Figure5}
\end{figure}

Experimental results in Figure \ref{fig:Figure5} demonstrate that the {4P4D (P-L20/D-H20)} configuration outperforms {4P4D (P-H20/D-L20)} with {34.67\%} faster average E2E on \verb|gov_report|, {40.1\%} on \verb|multi_news|, and {8.8\%} on \verb|qmsum|. Notably, vLLM's co-located prefill and decode phases exhibit significant throughput degradation during extended prefill operations, failing to meet TPOT constraints. Our heterogeneous deployment strategy effectively resolves this limitation. Compared to vLLM PD-colocated inference services, the {4P4D (P-L20/D-H20)} configuration achieves {48.9\%} faster E2E on \verb|gov_report|, {29.4\%} on \verb|multi_news|, and {15.2\%} on \verb|qmsum|. Concurrent TPOT improvements reach {44.57\%}, {24.2\%}, and {15\%} , respectively.

\subsection{\textbf{KV Cache transfer latency evaluation}}

In this section, we compare the KV Cache transfer latency based on simulated data for both single-server deployments (using a single L20 server) and multi-server heterogeneous deployments (using one L20 server and one H20 server) of Llama-3.1-8B-Instruct, with a deployment configuration of 1P1D. We also conduct a performance comparison with other disaggregated methods such as vLLM-Disagg and Mooncake. Additionally, we perform ablation experiments focused on transmission optimization for the NCCL pipeline, contrasting it with layer-wise transmission. The results are shown in the Table \ref{table:kvcache—latency}.

In single-server deployments, our NCCL pipeline reduces the average latency by 96.8\% compared to vLLM-Disagg, achieving a speedup of 31.5×. When compared to the RDMA-based backend employed in Mooncake, our approach decreases the average latency by 98.2\%, resulting in a speedup of 55.2×. In multi-server heterogeneous deployments, our NCCL pipeline reduces the average latency by 92\% compared to vLLM-Disagg and by 96.3\% compared to Mooncake, with corresponding speedups of 12.6× and 55.3×, respectively. Compared to the baseline layerwise-level KV Cache transfer, our method significantly reduces the number of NCCL kernel invocations per request, from 23,469 to just 1. In single-server deployments, the speedup is 24×, while in multi-server deployments, it is 15×.

%% file: latex/conclusion.tex
\section{Conclusion}

We propose FlowKV, an innovative PD-disaggregated LLM inference framework that supports both homogeneous and heterogeneous deployment. FlowKV introduces multiple KV cache transmission methods and achieves significant reductions in transmission latency when utilizing NCCL as the backend, demonstrating speedup ratios of 24× and 15× in single-node and multi-node configurations, respectively. Through Load-Aware Scheduler, FlowKV maximizes hardware utilization to enhance system throughput. Experimental results demonstrate a 25\% improvement in throughput compared to baseline systems, with significant advantages of 95\%, 40\%, and 35\% higher throughput over DistServe, Mooncake, and vLLM-Disagg, respectively. The framework exhibits scalable performance in heterogeneous GPU environments, delivering 15.2\%-48.9\% inference acceleration on  gov report, multi news, and qmsum datasets compared to baseline.

%% file: latex/Appendix.tex
\newpage 

\appendix


\section{Supplementary Experimental Results and Figures}



\begin{figure}[!ht]
    \centering
    \includegraphics[width=0.75\linewidth]{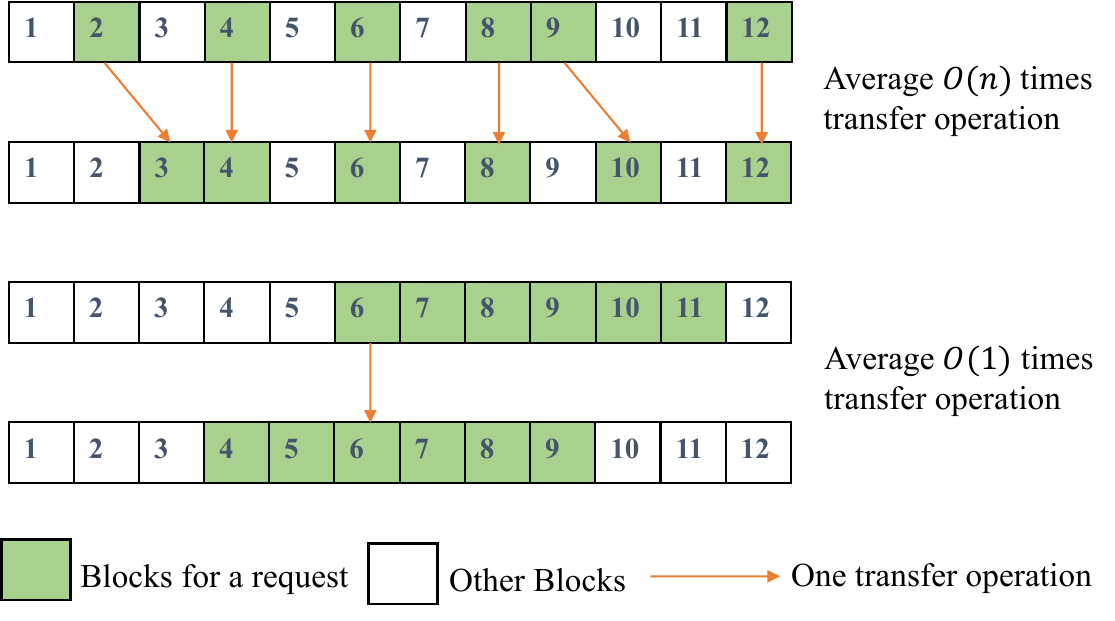}
    \caption{Comparison of the KV Cache transfer process in FlowKV with the pre-optimization approach. FlowKV aims to allocate the KV cache block IDs of both the sending and receiving instances within a contiguous memory segment as much as possible. Prior to the transfer operation, bidirectional segment alignment is performed to compare the block ID lists of both the sending and receiving instances, identifying N contiguous block IDs that are present on both sides. Since these N block IDs are memory-contiguous, they can be transferred in a single operation. The diagram illustrates an ideal scenario where the number of NCCL kernel calls in the KV Cache transfer operation is optimized from $O(n)$ times to $O(1)$.}
    \label{fig:figure3}
\end{figure}

\begin{table*}[!ht]
\centering
\renewcommand{\arraystretch}{1} 

\begin{tabular}{p{2.2cm}p{1.8cm}p{1.8cm}p{1.8cm}p{1.8cm}p{1.8cm}}
\toprule
\bf Setup & \bf Input/\newline Output & \bf Mooncake & \bf vLLM-Disagg & \bf FlowKV\newline Layerwise & \bf FlowKV \\
\midrule
Single  & 500/100 & 0.3010 & 0.1179 & 0.0678 & \bf0.0044 \\
Machine & 1000/100 & 0.5416 & 0.2314 & 0.1309 & \bf0.0075 \\
& 2000/100 & 1.0335 & 0.3435 & 0.2565 & \bf0.0126 \\
& 4000/100 & 1.3473 & 0.6670 & 0.5338 & \bf0.0236 \\
& 8000/100 & 2.0289 & 1.3382 & 1.1173 & \bf0.0447 \\
& 10000/100 & Failure & 1.7373 & 1.4121 & \bf0.0555 \\
& 12000/100 & Failure & 2.1894 & 1.7218 & \bf0.0681 \\
\midrule
Multiple & 500/100 & 0.3418 & 0.1197 & 0.1176 & \bf0.0080 \\
Heterogeneous& 1000/100 & 0.5820 & 0.1914 & 0.3262 & \bf0.0136 \\
Machines& 2000/100 & 0.8180 & 0.3444 & 0.4324 & \bf0.0260 \\
& 4000/100 & 1.4342 & 0.6681 & 0.8668 & \bf0.0519 \\
& 8000/100 & 2.1250 & 1.3462 & 1.6711 & \bf0.0993 \\
& 10000/100 & Failure & 1.7425 & 2.0719 & \bf0.1500 \\
& 12000/100 & Failure & 2.1974 & 2.4965 & \bf0.1759 \\
\bottomrule
\end{tabular}
    \caption{Comparison of KV Cache transfer latency based on the Llama-3.1-8B-Instruct model with a deployment configuration of 1P1D. The unit in the figure is seconds.}
    \label{table:kvcache—latency}
\end{table*}

\section{Load-Aware Scheduling }

\noindent\subsection{Application of Load-Aware Scheduling Scheme in Different Scenarios}

\label{Load-Aware-Application}

\textbf{Normal Load}. The global controller selects  the optimal \( P_t \) node for prefill requests with the goal of minimizing Time-To-First-Token (TTFT), taking into account KV-cache prefix hits and node loads. Subsequently,  \( D_t \) node are chosen for subsequent KV-cache transmission and decoding, with the goal of minimizing transmission latency.

\textbf{Imbalanced Load}. Due to the inherent delay between simple request routing and node workload adjustments, under computational imbalance scenarios, the global scheduler directly instructs idle nodes' hybrid schedulers to switch inference roles for several scheduling cycles, alleviating computational resource bubbles due to uneven loads.

\textbf{Extreme Load}. The global controller assesses load scores, and if their duration exceeds a threshold, it determines the need to scale up by increasing the number of nodes for specific roles, or conversely, scale down. After scaling, dynamic restructuring of the cluster is necessary. This strategy aids in maintaining high system availability and cost efficiency.

\noindent\subsection{Node Status Indicator Description}
We employ several metrics to evaluate node load status, including the running queue lengths ($L_{r}$), waiting queue lengths ($L_{w}$), swapped queue lengths ($L_{sw}$), and sending queue lengths ($L_{se}$) of prefill/decode requests, token budget ($T_{b}$), KV-cache utilization ($KV_{u}$), GPU utilization ($G_{u}$), and GPU memory bandwidth utilization ($MB_{u}$). The sending queue, a newly introduced component, is designed to manage requests that have completed the prefill phase and are now awaiting KV cache transmission to the decode node. Due to the bursty nature of GPU tasks (e.g., temporary high load followed by periods of idleness), instantaneous sampling can result in significant fluctuations. Therefore, we utilize a sliding-window approach to smooth out transient disruptions. After acquiring status indicators, we normalize all data to effectively assess each node's load status, and establish weight coefficients $w$ for various metrics based on task type, and compute a comprehensive load score ($C^{p}$ and $C^{d}$) through weighted calculations. These weight coefficients are determined through several successful experiments. The judgment of the entire system scenario is then derived from the load status of each node and by comparing it with predefined thresholds $\epsilon$.

\noindent\subsection{Load-Aware Scheduling scheme}
\begin{algorithm}
\caption{Load-Aware Scheduling Scheme}
\label{alg:Load-Aware}

\begin{algorithmic}[1]

\State \textbf{Input:} Inference requests $R$, weight coefficients $w$, predefined thresholds $\epsilon$
\State \textbf{Output:} Scheduling strategy

\While{system is running}
    \State Update the status of each node
    \For{each node $i$}
        \State $S_i \gets \left[
            \begin{array}{lll}
            L_r^{i, prefill}, & L_w^{i, prefill}, & L_{sw}^{i, prefill}, \\
            L_{se}^{i, prefill}, & L_r^{i, decode}, & L_w^{i, decode}, \\
            L_{sw}^{i, decode}, & L_{se}^{i, decode}, & T_{b}^i, \\
            KV_{u}^i, & G_{u}^i, & MB_{u}^i
            \end{array}
        \right]$
    \EndFor


    \State Calculate the comprehensive load score $C_i^{p}$ and $C_i^{d}$ for each node $i$
    \For{each node $i$}
        \State $C_i^{p} = w_{r, p} L_r^{i, prefill} + w_{w, p} L_w^{i, prefill} + w_{sw, p} L_{sw}^{i, prefill} + w_{se, p} L_{se}^{i, prefill} + w_{t, p} T_{b}^i + w_{kv, p} KV_{u}^i + w_{g, p} G_{u}^i + w_{mb, p} MB_{u}^i$
        \State $C_i^{d} = w_{r, d} L_r^{i, decode} + w_{w, d} L_w^{i, decode} + w_{sw, d} L_{sw}^{i, decode} + w_{se, d} L_{se}^{i, decode} + w_{t, d} T_{b}^i + w_{kv, d} KV_{u}^i + w_{g, d} G_{u}^i + w_{mb, d} MB_{u}^i$
    \EndFor
    \State {Calculate the comprehensive load scores $C^{p}$ and $C^{d}$:}
    \State $C^{p} = \frac{1}{N} \sum_i C_i^{p}$ for prefill node
    \State $C^{d} = \frac{1}{M} \sum_i C_i^{d}$ for decode node
    \State Determine the scenario based on $C^{p}$ and $C^{d}$
    
    \If {$|C^{p}| \leq  \epsilon_{p}^{low} \ \text{and} \ |C^{d}| \leq \epsilon_{d}^{low} $ \ \ \text{(normal load)}}
        \State \text{Schedule prefill request:} 
            \State  Among all options \(P_i\) in the set \(\mathcal{P}\), select the option \(P_t\) that minimizes the time to first token (TTFT), subject to a cache prefix hit condition on \(P_i\) and given the state \(S_i\)
        \State \textbf{Forward} $R$ to $P_{t}$ 
        \State \text{Schedule decode request:} 
            \State  Choose from options \(D_i\) in the set \(\mathcal{D}\), the option \(D_t\) that minimizes the transmission latency from already selected prefill option \(P_t\) 
        \State \textbf{Forward} $R$ to $D_{t}$ 
    \pagebreak
    \ElsIf {$|C^{p}| \leq \epsilon_{p}^{high} \ \text{and} \ \ |C^{d}| \leq \epsilon_{d}^{high}$ \ \ \text{(imbalanced load)}}
        \State $idleNodes \gets \{i | C_i < \text{threshold}\}$
        \State Notify local scheduler of $idleNodes$ to adjust the priorities of the hybrid scheduler. 
        \State Cycle scheduling until global controller sends \textbf{terminate\_signal}  
    \Else
        \State Increase/decrease number of instances/nodes
        \State Perform dynamic reconfiguration of the cluster
    \EndIf
\EndWhile
\end{algorithmic}
\end{algorithm}